# Computational Investigations of the Lithium Superoxide Dimer Rearrangement on Noisy Quantum Devices

Qi Gao,[1,2] Hajime Nakamura,[2,3] Tanvi P. Gujarati,[4] Gavin O. Jones,[4] Julia E. Rice,[4] Stephen P. Wood,[5] Marco Pistoia,[5] Jeannette M. Garcia[4] and Naoki Yamamoto[2]

[1]Mitsubishi Chemical Corp.
[2]Quantum Computing Center, Keio University
[3]IBM Research – Tokyo
[4]IBM Research – Almaden
[5]IBM Thomas J. Watson Research Center

Currently available noisy intermediate-scale quantum (NISQ) devices are limited by the number of qubits that can be used for quantum chemistry calculations on molecules. We show herein that the number of qubits required for simulations on a quantum computer can be reduced by limiting the number of orbitals in the active space. Thus, we have utilized ansätze that approximate exact classical matrix eigenvalue decomposition methods (Full Configuration Interaction). Such methods are appropriate for computations with the Variational Quantum Eigensolver algorithm to perform computational investigations on the rearrangement of the lithium superoxide dimer with both quantum simulators and quantum devices. These results demonstrate that, even with a limited orbital active space, quantum simulators are capable of obtaining energy values that are similar to the exact ones. However, calculations on quantum hardware underestimate energies even after the application of readout error mitigation.

## INTRODUCTION

Quantum computing is a method of computation that possesses the potential to surpass conventional computing (so-called classical computing). While the theoretical framework governing quantum computing has been established for decades, and algorithms have been developed for a variety of application areas, the field has recently experienced a surge in interest due to newly demonstrated success in manufacturing qubit devices. However, device technology is still in its infancy, and the quantum devices currently in operation, known as Noisy Intermediate-Scale Quantum (NISQ)[1] devices, depend on hybrid approaches involving the use of qubits in combination with classical computing architectures.

Quantum computing possesses enormous near-term potential for the development of applications in a number of areas, including quantum chemistry, for which finding eigenvalues of eigenvectors is an intractable problem for classical computers. Quantum computing may be particularly effective for such problems, and algorithms such as Quantum Phase Estimation (QPE)[2] and the Variational Quantum Eigensolver (VQE)[3] have been developed to find eigenvalues for approximate, but highly accurate, solutions to the Schrödinger equation. VQE in particular, due to a comparatively shorter circuit than QPE, reduces time requirements for qubits to remain coherent, and has been effectively utilized to perform quantum chemistry calculations on NISQ devices.[4]

IBM researchers have demonstrated the use of VQE in combination with heuristic trial wavefunctions designed for state preparation in investigations of the ground state dissociation profiles of hydrides on a quantum device.[4] Those calculations demonstrated that energies computed for dissociation profiles on quantum devices are nearly similar to those computed with a classical matrix eigenvalue decomposition method (Full Configuration Interaction, or Full CI, or FCI) for the hydrogen molecule, but profiles deviate significantly from chemical accuracy for distances far from the equilibrium geometries of lithium hydride and beryllium hydride.

Those pioneering investigations inspired us to evaluate the performance of quantum computers in determining reaction energetics for species near equilibrium distances in order to predict reaction profiles and mechanisms of reactions. Such investigations possess enormous implications for the ability to develop new types of reaction methodologies, as well as to design new catalysts and new materials.

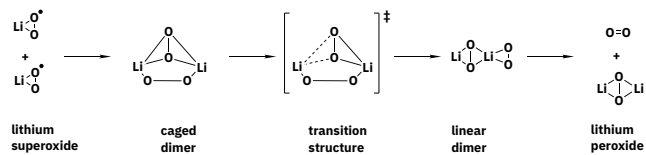

**Fig. 1**. Formation of lithium peroxide and molecular oxygen via the rearrangement of caged lithium superoxide dimer into linear superoxide dimer.

This manuscript describes our investigations on the dimerization of lithium superoxide, an interesting test case for these types of calculations that could be applied to a problem with potential real-world implications. The motivation for these investigations derives from the recent interest in developing lithium-air (Li/$O_2$) batteries. These are promising electrochemical cells and may possess higher energy densities than widely-used lithium ion batteries.[5–9] During discharge, lithium combines with superoxide, obtained from the reduction of oxygen at the cathode, to produce lithium peroxide ($Li_2O_2$) via a lithium superoxide ($LiO_2^{\bullet}$) intermediate.[10–12] Interestingly, though, in addition to forming lithium peroxide alone, lithium superoxide may also dimerize to generate lithium peroxide plus an oxygen molecule via a rearrangement process (Fig. 1).[11,12] The overall effect of these processes is that molecular oxygen is consumed at the lithium cathode during discharge to produce lithium peroxide and regenerate residual oxygen gas.

Quantum chemistry investigations have previously been performed to predict the structures, mechanisms and energies for the production of lithium peroxide and oxygen from lithium superoxide dimerization.[11,12] Those investigations utilized geometries optimized with the Becke, 3-parameter, Lee–Yang–Parr density functional theory (B3LYP DFT) method to obtain Coupled Cluster Singles and Doubles with perturbated Triples (CCSDT) energies after extrapolation to the Complete Basis Set (CBS) limit.[11] Critically, investigations of potential mechanisms found that the free energy surface for rearrangement is rather flat, and many transition structures could potentially lead to formation of lithium peroxide and oxygen from different geometries of the lithium superoxide dimer.

While there are many possible pathways due to the flat nature of the free energy surface,[11] in this manuscript we will focus on the pathway involving rearrangement of the caged lithium superoxide dimer into the linear lithium superoxide dimer (henceforth referred to as the *reactant* and *product*, respectively), which can be viewed as a complex comprising the oxygen molecule bound to lithium peroxide. The free energy barrier for this reaction at 298K was previously predicted to be 17 *mHa*, and the linear dimer is comparatively less stable than the caged dimer.[11,12] The overall formation of completely separated lithium peroxide and oxygen from the caged dimer was also predicted to be endergonic.[11,12]

In order to use a NISQ device to perform quantum chemistry calculations on molecules such as those of interest to this manuscript, one would have to ensure the availability of the number of logical qubits depending on the problem of interest. The number of qubits (which map directly to the number of spin orbitals of molecules) required for this problem is 60 for the full set of atomic orbitals with a minimal basis set, or 48, if core orbitals, which may be reliably neglected since they do not interact with valence orbitals, are frozen. At present however, we are limited in the number of qubits we can reliably use for computation. Consequently, further qubit reductions would have to be employed in order to make use of such NISQ devices for investigations such as the one that we focus on in this manuscript. We demonstrate that the reduction of orbitals to just the highest occupied molecular orbitals (HOMO) and lowest unoccupied molecular orbitals (LUMO) of the stationary points can effectively reduce this problem down to two qubits with a 6-31G(d,p) basis set[13–20] for the investigation of the complete mechanism of this rearrangement reaction.

**METHODOLOGY**

The reaction under investigation involves conversion of the caged lithium superoxide dimer into the linear superoxide dimer via a transition structure containing partially broken bonds in the "bridge" formed by lithium and oxygen atoms. The overall strategy for these investigations involved initial preprocessing with classical quantum chemistry codes on conventional computers to generate optimized geometries and guess orbitals prior to performing computations with quantum simulators or devices.

Classical calculations were performed with the Jaguar module[21] contained in the Schrödinger software suite,[22] in which initial molecular geometries of the reactant and product from previously published literature sources[11] were used and then optimized with the B3LYP[23–26] method, to which dispersion corrections, as described by Grimme et al.,[27] were applied. The resulting B3LYP-D3 procedure was coupled with the 6-311++G(d,p) basis set.[28,29] An initial guess for the transition structure connecting the reactant and product dimers was determined by using Auto TS,[30] a module of the Schrödinger software package. Because of the flat nature of the potential energy surface (PES) in the region of the transition structure,[11] an analytic Hessian was used for the TS search and for optimization of the initial-guess geometry. Intrinsic reaction coordinate (IRC) analysis was performed on the optimized geometry of the transition structure in order to confirm that it was connected to the caged and linear lithium superoxide dimers.

In principle, one should be able to perform a Full CI calculation more efficiently on a quantum computer than on a classical computer. However, only a few molecular orbitals can currently be simulated on such devices due to the fact that the number of qubits is still relatively small, the devices are noisy, and error correction has not yet been fully realized. To address this issue, a set of molecular orbitals comprising the HOMO and LUMO orbitals for the reactant, product and TS were selected because these molecular orbitals provide the largest contributions of electron correlation to the energy of the system.

Having selected the set of orbitals belonging to the active space for the stationary points, quantum computations were performed with quantum simulators and devices using VQE.[3] Note that, in contrast with quantum chemistry on conventional computers, in which the molecular spin state needs to be predefined in order to compute the most stable electronic state, the VQE algorithm automatically minimizes the energy to the most stable spin state because the direction of the spin freely rotates at every qubit.[31] An addendum to this is that an initial state needs to be precomputed via a classical algorithm on a conventional computer. The HF singlet state has been chosen as the initial state for all of the calculations described in this manuscript, because previous publications have shown that this is a good choice for an initial state.[32] For consistency with the active space that we have chosen for these investigations, the energies of orbitals from the singlet state were also used for the orbitals that remained inactive. Since previous publications have indicated that the triplet surface for the rearrangement of the lithium superoxide dimer is more stable than that of the rearrangement on the singlet surface,[11] we recognize that using energies of the singlet spin states for the inactive orbitals may cause VQE to overestimate the energies of singlet spin states in the qubit space. Consequently, we intend to investigate results obtained with VQE calculations with different starting spin states in a future publication.

The Aqua module contained in Qiskit version 0.10[33] with an interface to PySCF[34] was used for all VQE calculations. The statevector simulator contained in the Aer mode of Qiskit was used for all simulations. The Conjugate Gradient (CG) method[35] for energy minimization was used for calculations on simulators and the Simultaneous Perturbation Stochastic Approximation (SPSA)[36,37] method was used for calculations on quantum devices.

A variety of trial wavefunctions for these orbitals were also used on simulators and quantum devices. Thus, the Unitary Coupled Cluster Singles and Doubles (UCCSD) method,[38–40] which provides energies that are similar to those provided by the Exact Eigensolver (Full CI or the exact diagonalization method supplied by Qiskit Aqua for computing reference values) and preserves the number of particles in the calculation, was compared with three heuristic variational forms (Ry, RyRz and SwapRz)[41] with a circuit depth of 1 for calculations run on a quantum simulator. Calculations performed on quantum devices utilized the Ry variational form. The Ry and RyRz variational forms can be used with the parity mapping transformation[42] for transforming the fermionic Hamiltonian into the qubit Hamiltonian. Because of the symmetry properties of parity mapping, the total number of required qubits can be reduced by two without loss of precision whenever parity mapping is an option. On the other hand, SwapRz preserves the number of particles when used with the Jordan-Wigner transformation,[43] though this mapping does not have the symmetry properties of parity mapping, and application of two-qubit reduction is therefore not possible. The HF method combined with the 6-31g(d,p) basis set was chosen as the initial state for the trial wavefunctions.

The IBM 20-qubit machine, ibmq_poughkeepsie, was used to perform experiments on quantum devices. A set of two qubits with direct connectivity, small readout error rates, and small two-qubit error rates was chosen for those experiments. 8192 shots were used to measure the expectation value of each Pauli term in the Hamiltonian. Since current quantum devices are error prone, the readout error mitigation technique [20] (which is supported by Qiskit) was used to improve upon errors sustained during qubit readout. The measurement calibration matrix was updated every 30 minutes to ensure that experimental conditions during readout measurement calibration and VQE iteration were similar. The final energy values reported for VQE experiments on the hardware are based on the lowest moving average over 100 VQE data points, with a moving window of 10 data points. SPSA optimizations were carried out for a total of 500 iterations for the reactant, product and the TS.

## RESULTS AND DISCUSSION

*1. Quantum chemistry calculations on a classical computer.*
HOMO and LUMO orbitals were selected to describe the active orbital space for the reactant, transition structure and product. The set of chosen orbitals for all three stationary points include the p-type orbitals on the oxygen atoms.

The energies of the reactant, TS and product obtained with HF, UHF-CCSD (Unrestricted Hartree-Fock CCSD) and FCI using the chosen active space are shown in Table 1 and the electronic energy profile is shown in Fig. 2. When UHF is used, the reactant is much higher in energy than the TS and product suggesting that the reaction mechanism possesses no barrier; this is presumably due to the fact that HF does not adequately account for electron correlation. In contrast, the correlation energy of the reactant is much larger than that of the TS and product when UHF-CCSD or FCI are used. As a consequence, the energy of the reactant becomes much more stable than that of the TS and the energetic barrier becomes 83 *mHa* and 82 *mHa*, respectively, for UHF-CCSD and FCI.

**Table 1.** HF, UHF-CCSD and FCI energies, in hartrees, of the reactant, TS and product computed with the 6-31G(d,p) basis set. Energies relative to the reactant energies are shown in parentheses. Calculations involving UHF-CCSD and FCI were performed on active spaces containing the HOMO and LUMO. Note that for the product, the active space with the HOMO and LUMO does not capture the effect of the important excitations. (1 *Ha* = 627.5095 kcal/mol)

|  | HF | UHF-CCSD | FCI |
| --- | --- | --- | --- |
| *Reactant* | -314.063494 (0.000) | -314.182438 (0.000) | -314.182440 (0.000) |
| *Transition state* | -314.096974 (-0.033) | -314.100268 (0.082) | -314.100714 (0.082) |
| *Product* | -314.119501 (-0.056) | -314.119501 (0.062) | -314.119501 (0.063) |

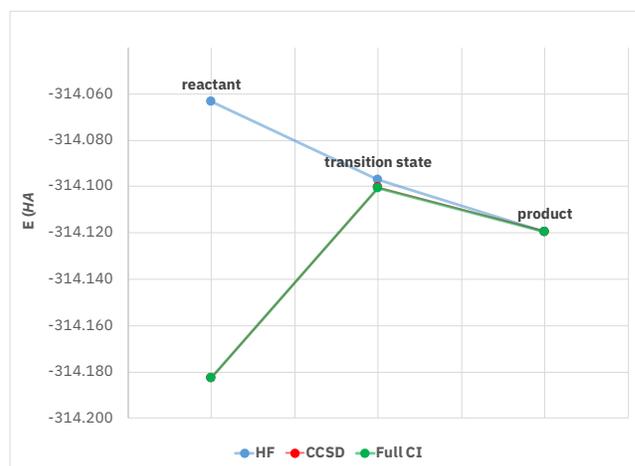

**Fig. 2**. Electronic energy profiles obtained with the HF, UHF-CCSD and FCI methods with the 6-31G(d,p) basis set for the HOMO and LUMO active orbitals of the reactant, transition state and product. (1 *Ha* = 627.5095 kcal/mol)

**Table 2.** Relative energies (in millihartrees) of triplet states computed with B3LYP-D3, UHF-CCSD and UHF-CCSD(T) for the HOMO and LUMO active orbitals of the reactant, TS and product. All calculations were performed with the 6-311++G(d,p) basis set and utilized structures optimized with B3LYP/6-311++G(d,p). (1 *mHa* = 0.6275095 kcal/mol)

|  | B3LYP-D3 | UHF-CCSD | UHF-CCSD(T) |
| --- | --- | --- | --- |
| *Reactant* | 0 | 0 | 0 |
| *Transition State* | 18 | 43 | 40 |
| *Product* | 17 | 37 | 39 |

In order to compare the results of this study with previously published results, we have computed the energies of the stationary points of the reaction with B3LYP-D3, UHF-CCSD and UHF-CCSD(T) using the 6-311++G(d,p) basis set and collected the results in Table 2. The energy barriers for calculations involving UHF-CCSD and UHF-CCSD(T) are 25 *mHa* and 22 mHa higher, respectively, than that found for B3LYP-D3. These results are similar to results previously reported in the scientific literature[11] which indicated that B3LYP underestimates the barriers and thermodynamics of lithium superoxide dimer rearrangement in comparison to UHF-CCSD(T). Notably, the energy barrier predicted with UHF-CCSD(T) is about 12 *mHa* larger than previously predicted,[11] which may signify that the transition state geometry is different from that which was previously found.

With the geometries and energies from the classical computation in hand, we then turned to computations with a quantum simulator to determine energies and reaction profiles for the HOMO/LUMO active spaces of these stationary points. Fig. 3 shows comparisons of energies of the reactant, TS and product computed with UCCSD and heuristic trial wavefunctions (Ry, RyRz and SwapRz) with the exact eigensolver at circuit depth = 1 with the 6-31G(d,p) basis set. These results demonstrate that UCCSD and all of the heuristic ansatzes predict similar energies to the exact eigensolver. It is notable, though, that FCI/6-31G(d,p) predicts that the energetic barrier for the rearrangement is 81 *mHa*, which is 41 *mHa* larger than the barrier predicted by UHF-CCSD(T)/6-311++G(d,p),[11,12] which is most likely due to the comparatively smaller size of this basis set.

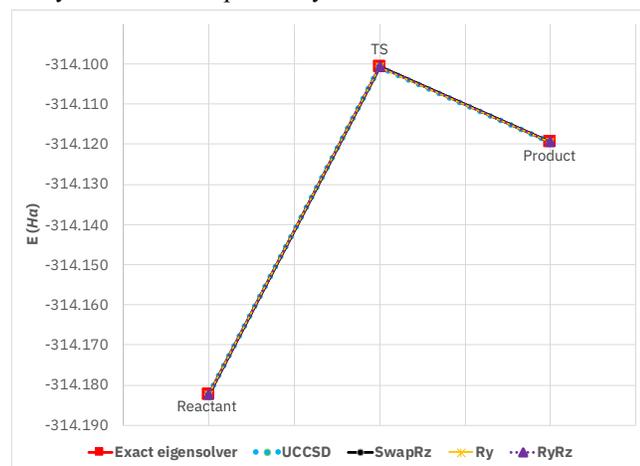

**Fig. 3**. Energies of reactant, transition state and product as computed by the (a) UCCSD (b) SwapRz (c) Ry and (d) RyRz ansatzes. (1 *Ha* = 627.5095 kcal/mol)

The number of CNOT gates in the circuit, and the number of optimization parameters used in the VQE algorithm for all the heuristic methods have been compared in Table 4 in order to determine which trial wavefunction would be suitable for use on the real device. As these results show, when depth = 1, circuits involving the use of UCCSD and SwapRz wavefunctions possess 4 CNOT gates, whereas only 1 CNOT gate is required for Ry and RyRz wavefunctions. Moreover, increasing the depth to 2 results in the addition of 4 CNOT gates when the UCCSD and SwapRz ansatzes are used, but a similar increase in the depth only increases the number of CNOT gates required for Ry and RyRz wavefunctions by a single gate. Similarly, the number of optimization parameters required for VQE calculations linearly increases in the order UCCSD < Ry < SwapRz < RyRz for circuit depth equaling 1, but at higher circuit depths the order for the increase in the number of optimization parameters changes to Ry < UCCSD ≈ SwapRz < RyRz.

In order to obtain reliable results from a real device in a short amount of time one would need a wavefunction that has few CNOT gates, since the CNOT error rate influences the accuracy of the computed results, and also few optimization parameters, since the number of optimization parameters is proportional to the number of iterations required for energy convergence. Accordingly, because Ry possesses comparatively fewer CNOT gates and optimization parameters than the other wavefunctions even with higher circuit depths, it was selected as the most suitable ideal wavefunction of those available for calculations on the quantum device.

**Table 3.** Number of CNOT gates and optimization parameters required at circuit depths, d, ranging from 1-3 with the UCCSD, SwapRz Ry, and RyRz ansatzes.

|  | Number of CNOT gates | | | Number of Optimization Parameters | | |
|---|---|---|---|---|---|---|
|  | d=1 | d=2 | d=3 | d=1 | d=2 | d=3 |
| *UCCSD* | 4 | 8 | 12 | 3 | 6 | 9 |
| *SwapRz* | 4 | 8 | 12 | 5 | 8 | 11 |
| *Ry* | 1 | 2 | 3 | 4 | 6 | 8 |
| *RyRz* | 1 | 2 | 3 | 8 | 12 | 16 |

2. *Quantum chemistry simulations on quantum devices.* Based on results obtained from the quantum simulator experiments, the Ry trial wavefunction with circuit depth = 1 was used as the trial wavefunction and readout error mitigation was applied to raw measurement counts obtained from calculations on the quantum devices.

Fig. 4a-c shows the trend in energy minimization for the reactant, TS and product, respectively, as a function of the VQE iterations for the readout error mitigated and unmitigated cases. Notably, only the reactant (Fig. 4a) exhibits a large offset energy in the VQE iterations during energy minimization. This is presumably due to incomplete convergence of the SPSA algorithm to the lowest possible energy state in an energy landscape with other proximal local minima or alternatively the ground state may not be well described by the Hartree-Fock wavefunction.

Fig. 4d summarizes the results in Table 4 which shows that the results obtained from the quantum hardware without error mitigation overestimate values derived from the exact eigensolver by 14-33 *mHa*.

Overall, with readout error mitigation included, the results still overestimate the exact values by 11 *mHa* and 2 *mHa* for the reactant and product, respectively, after 500 VQE iterations. In contrast, error mitigation leaves the energy for the transition state almost unaffected by running on the quantum device. As a consequence, barriers predicted by computations performed on the ibmq_poughkeepsie quantum hardware underestimate the exact values by 18 *mHa* and 14 *mHa* for calculations performed without and with readout mitigation, respectively.

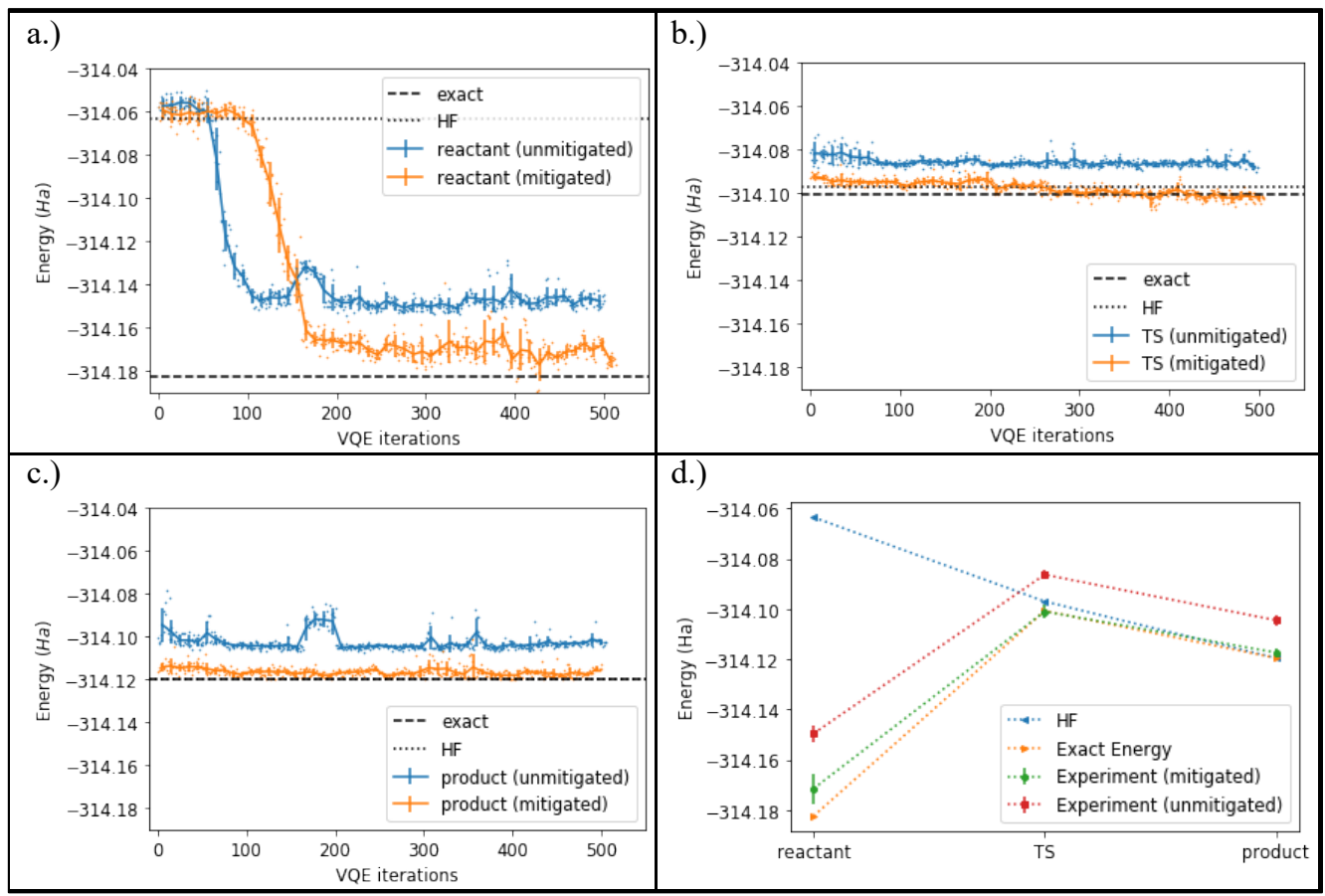

**Fig. 4.** VQE iterations for calculations on the ibmq_poughkeepsie quantum device for the HOMO/LUMO active spaces of a.) reactant b.) TS and c.) product of the lithium superoxide caged dimer d.) Summary of the results obtained for the three structures through hardware experiments. (1 *Ha* = 627.5095 kcal/mol)

**Table 4.** Comparison of energies (in hartrees) computed on the HOMO and LUMO active space for reactant, transition state and product with the exact eigensolver and the Ry heuristic ansatz on simulators and on the ibmq_poughkeepsie 20 qubit quantum device using the Ry ansatz at depth = 1 with and without readout error mitigation. Energy values with reference to reactant energies, in millihartrees, are shown in parentheses. (1 mHa = 0.6275095 kcal/mol)

|  | HF | Ry, simulator | Ry, without error mitigation | Ry, error mitigated experiment | Exact |
|---|---|---|---|---|---|
| *Reactant* | -314.0634943 (0) | -314.182440 (0) | -314.149589 ± 0.003158 (0) | -314.171507 ±0.006034 (0) | -314.182440 (0) |
| *TS* | -314.096974 (-33) | -314.100714 (82) | -314.086306 ± 0.001679 (63) | -314.100982 ±0.002659 (67) | -314.100715 (81) |
| *Product* | -314.119501 (-56) | -314.119501 (63) | -314.104445 ± 0.002146 (45) | -314. 117326 ±0.001868 (54) | -314.119501 (63) |

## CONCLUSIONS

In summation, the VQE algorithm has been used to investigate the rearrangement of the lithium superoxide dimer from its caged structure into its linear analogue with a variety of ansatzes on a quantum simulator and on NISQ hardware. This study utilized a reduced active space comprising only the HOMO and LUMO orbitals of the reactant, transition state and product. This choice resulted in a good description of the reactant and transition state and also served to reduce the number of qubits for the computation.

Calculations performed with a quantum simulator using this limited active space reproduced exact values derived from full configuration interaction and a classical matrix eigenvalue decomposition method. In contrast, computations with the ibmq_poughkeepsie quantum device overestimate the exact values and overestimates energetics. Inclusion of error mitigation reduces the amount of overestimation for the energies of the stationary points. In particular, transition state energies, were almost exactly replicated after error mitigation was applied, but reactant and product energies were still overestimated. The overall effect of this is that the energy barrier for the rearrangement was underestimated by performing calculations on quantum hardware and including error mitigation. Nonetheless, we believe that further improvements can be made to these results by including techniques that additionally correct for incoherent gate errors.[44]